\documentclass[iop]{emulateapj}
\usepackage{subfigure}
\usepackage{xcolor}
\usepackage{graphicx}
\usepackage{xspace}
\usepackage{amsmath, amsthm, amssymb}
\shorttitle{Multiple Hot Jupiter Populations}
\shortauthors{Nelson et al.}

\newcommand{\ms}{\ensuremath{\textup{m\,s}^{-1}}\xspace}
\newcommand{\aroche}{\ensuremath{a_{\rm Roche}}\xspace}

\newcommand{\FordRasio}{\textup{FR06}\xspace}

\newcommand{\exoplanetsorg}{{\tt exoplanets.org}\xspace}
\newcommand{\Stan}{{\tt Stan}\xspace}

\begin{document}

\title{Evidence for Two Hot Jupiter Formation Paths }

\author{Benjamin E. Nelson\altaffilmark{1,2}, 
Eric B. Ford \altaffilmark{3,4},
Frederic A. Rasio \altaffilmark{1}}

\altaffiltext{1}{Center for Interdisciplinary Exploration and Research in Astrophysics (CIERA) and Department of Physics and Astronomy, Northwestern University, 2145 Sheridan Road, Evanston, IL 60208, USA}
\altaffiltext{2}{Northwestern Institute for Complex Systems, 600 Foster Street, Evanston, IL 60208, USA}
\altaffiltext{3}{Center for Exoplanets and Habitable Worlds, The Pennsylvania State University, 525 Davey Laboratory, University Park, PA, 16802, USA}
\altaffiltext{4}{Department of Astronomy \& Astrophysics, The Pennsylvania State  University, 525 Davey Laboratory, University Park, PA 16802, USA}

\email{email: benelson@northwestern.edu} 

\begin{abstract}
Disk migration and high-eccentricity migration are two well-studied theories to explain the formation of hot Jupiters.
The former predicts that these planets can migrate up until the planet-star Roche separation (\aroche) and the latter predicts they will tidally circularize at a minimum distance of 2\aroche.
Considering long-running radial velocity and transit surveys have identified a couple hundred hot Jupiters to date, we can revisit the classic question of hot Jupiter formation in a data-driven manner.
We approach this problem using data from several exoplanet surveys (radial velocity, Kepler, HAT, and WASP) allowing for either a single population or a mixture of populations associated with these formation channels, and applying a hierarchical Bayesian mixture model of truncated power laws of the form $x^{\gamma-1}$ to constrain the population-level parameters of interest (e.g., location of inner edges, $\gamma$, mixture fractions).
Within the limitations of our chosen models, we find the current radial velocity and Kepler sample of hot Jupiters can be well explained with a single truncated power law distribution with a lower cutoff near 2\aroche, a result that still holds after a decade, and $\gamma=-0.51\pm^{0.19}_{0.20}$.
However, the HAT and WASP data show evidence for multiple populations (Bayes factor $\approx 10^{21}$).
We find that $15\pm^{9}_{6}\%$ reside in a component consistent with disk migration ($\gamma=-0.04\pm^{0.53}_{1.27}$) and $85\pm^{6}_{9}\%$ in one consistent with high-eccentricity migration ($\gamma=-1.38\pm^{0.32}_{0.47}$).
We find no immediately strong connections with some observed host star properties and speculate on how future exoplanet surveys could improve upon hot Jupiter population inference.

\end{abstract}

\section{Introduction}
More than two decades after their discovery, hot Jupiters remain elusive in regard to their primary formation channel.
They likely formed at several AU, beyond their host stars' ice lines \citep{Mizuno80, Pollack96}.
Theories for bringing these giants into sub-Mercury orbits can generally be lumped into two categories.
The first is a gentle, slow orbital decay through interactions with a protoplanetary disk \citep{GoldreichTremaine80, Lin96, Ward97, Murray98}.
The second is tidal circularization of a planet on a highly eccentric orbit, created from sudden or secular gravitational encounters with other massive bodies \citep{RasioFord96, WuMurray03, FabryckyTremaine07, WuLithwick11, Naoz11}.

Very simple theories of disk versus eccentric migration also make testable predictions.
Previous studies have investigated these theories in several contexts, including but not limited to spin-orbit misalignment \citep{Naoz12, Rogers12}, stellar metallicity \citep{Dawson13}, and the presence of additional companions \citep{Knutson14, Ngo15, Piskorz15, Bryan16, Ngo16, SchlaufmanWinn16}.
The focus of this paper will be in the context of the Roche separation \aroche between the planet and host star.
For an infinitely compressible planet, this is given by $\aroche = 2.16 R \mu^{-1/3}$, where $R$ is the physical radius of the planet and $\mu$ is the planet-star mass ratio \citep{Faber05}.

In disk migration, planets can inwardly migrate up until \aroche.
A Jupiter-sized planet may be consequently stripped of its gaseous envelope or tidally disrupted depending on the details of mass transfer at this distance \citep{Valsecchi14, Valsecchi15}.
In eccentric migration, \aroche is the minimum periastron distance for the planet to survive without considerable mass loss.
Such an orbit would circularize at a semi-major axis greater than $2\aroche$.
Pioneering radial velocity surveys showed the observed distribution of hot Jupiter semi-major axes normalized to their respective Roche separations ($x\equiv a/\aroche$) has an inner cutoff near 2 \citep[\FordRasio henceforth]{FordRasio06}, supporting the prediction from an eccentric migration history.
At the time, the sample size was 21 planets, most with only minimum mass ($M\sin{i}$) measurements.

One decade since then, there are many more radial velocity planets as well as dedicated photometric transits surveys that are able to probe physical radii.
Stars targeted by ground-based transit surveys are often amenable to radial velocity follow-up, so a physical mass (rather than $M\sin{i}$) can be inferred for these planets, providing individual $x$ values.

By combining planet discoveries from various surveys into an updated sample, we may be able to say something more precise about the population statistics of hot Jupiters with respect to \aroche and how they compare to predictions of well-studied formation theories.
Rather than motivating and testing a particular theory to compare to the observations (e.g., through planet population synthesis models), we approach this problem from a data-driven perspective.
In particular, could we infer a single population or disentangle two populations from the data, where each population is consistent with a disk migration and/or eccentric migration history?

In this paper, we show how a hierarchical Bayesian framework can address this classic question of hot Jupiter formation.
In \S \ref{sec-data}, we describe the datasets used for the study.
In \S \ref{sec-statmodel}, we describe the statistical models applied to each dataset.
In \S \ref{sec-results}, we show the results of our inference and discuss the details and caveats in \S \ref{sec-discussion}.

\section{Data}
\label{sec-data}
\begin{deluxetable}{cc}
\tablecaption{Notation and descriptions of selected variables used in this paper.\label{tbl-ref} }
\tablewidth{0pt}
\tablehead{ \colhead{Parameter} & \colhead{Description} }
\startdata
\hline
$\mathcal{M}$ & $\textup{models}$ \\
\hline
$\mathcal{M}_{1}$ & $\textup{1-component truncated Pareto}$ \\
$\mathcal{M}_{1,\sigma}$ & $\mathcal{M}_1 \textup{ with observational uncertainties}$ \\
$\mathcal{M}_{2}$ & $\textup{2-component truncated Pareto mixture}$ \\
$\mathcal{M}_{2,\sigma}$ & $\mathcal{M}_2 \textup{ with observational uncertainties}$ \\
\\
\hline
$\vec{d}$ & $\textup{data}$ \\
\hline
$\hat{P}$ & $\textup{observed orbital period}$ \\
$\hat{M}$ & $\textup{observed planet mass}$ \\
$\hat{R}$ & $\textup{observed physical radius}$ \\
$\sigma_{\hat{P}}$ & $\textup{uncertainty in observed orbital period}$ \\
$\sigma_{\hat{M}}$ & $\textup{uncertainty in observed planet mass}$ \\
$\sigma_{\hat{R}}$ & $\textup{uncertainty in observed physical radius}$ \\
\\
\hline
$\vec{\alpha}$ & $\textup{individual parameters}$ \\
\hline
$i$ & $\textup{true inclination of orbital plane to sky plane}$ \\
$P$ & $\textup{true orbital period}$ \\
$M$ & $\textup{true planet mass}$ \\
$R$ & $\textup{true physical radius}$ \\
$x$ & $\textup{planet semi-major axis divided by Roche separation}$ \\
$x_{edge}$ & $x \textup{ assuming true mass = minimum mass}$ \\
\\
\hline
$\vec{\beta}$ & $\textup{hyper-parameters}$ \\
\hline
$x_l$ & $\textup{lower limit of} x \textup{ for one-component distribution}$  \\
$x_{l,1}$ & $\textup{lower limit of } x \textup{ for first majorixture component}$  \\
$x_{l,2}$ & $\textup{lower limit of } x \textup{ for second mixture component}$  \\
$x_u$ & $\textup{upper limit of } x$  \\
$\gamma$ & $\textup{power law index for one-component distribution in } x$ \\
$\gamma_1$ & $\textup{power law index for first mixture component in } x$ \\
$\gamma_2$ & $\textup{power law index for second mixture component in } x$ \\
$f_i$ & $\textup{ith mixture component fraction}$ \\
\enddata
\end{deluxetable}

For this analysis, we consider four different samples of exoplanets discovered from various surveys: radial velocity, HATNet \citep{Bakos04}, WASP \citep{Street03}, and Kepler \citep{Santerne16}.
For all of these datasets, we restrict the orbital periods to be $P < 30 \textup{d}$ and velocity semi-amplitude $K > 30 \ms$ to minimize selection effects \citep[\FordRasio]{Cumming04}.
We recognize that these transit surveys are prone to observational biases and selection effects not present in long-running RV surveys, which we will address later in this section.

There are multiple well-known exoplanet databases to choose from and each have their own strengths.
The NASA Exoplanet Archive \citep{Akeson13} consistently keeps up with new planet discoveries and refinements on orbital elements but is therefore likely to contain some controversial planet candidates.
\exoplanetsorg \citep{Han14} is slower to update planet discoveries and properties but is much more thorough in vetting planet claims.
The choice here comes down to an incomplete and slightly out-of-date list of planets versus an uncurated but up-to-date list.
We ultimately analyze planets queried from both databases but present the Exoplanet Archive results in this manuscript.
We will describe any difference datasets and results in Section \ref{discussion-data} and Section \ref{discussion-results} respectively.

We queried the Exoplanet Archive on March 4, 2017 to obtain the orbital properties for these planets given the above period and semi-amplitude restrictions above.
For the radial velocity sample, we only include planets discovered through the radial velocity method, stressing that this set does not include planets discovered by transit surveys that had follow-up radial velocity measurements.
For the HAT and WASP samples, we filter for any planet with a WASP or HAT substring in its name and ensure they were discovered via the transit method.
The only exception was WASP-94, a wide binary system with both stars hosting their own planets \citep{Neveu14}.
The hot Jupiter around the secondary star was detected through radial velocity and is thus excluded from our sample.

For those planets without measured radii, we assume a radius of $1.2\pm0.1 R_J$, which is roughly how known exo-Jupiter radii are distributed.
For the Kepler sample, we consult the planets with ``secured'' planetary status listed in Table 7 of \citet{Santerne16} and apply the same period and semi-amplitude cutoffs.
No planet was seen in multiple datasets.

Now $x$ can be rewritten in terms of radius ($R$), orbital period ($P$), and planet mass ($M$) as such:
\begin{equation}
x\equiv \frac{a}{\aroche} = 0.462 \left(\frac{P}{1\,\rm yr}\right)^{2/3} \left(\frac{M}{1\,M_\odot}\right)^{1/3} \left(\frac{R}{1\,\rm AU}\right)^{-1}.
\end{equation}
We will consider two models, one that neglects and one that incorporates observational uncertainties in these parameters (see Section \ref{sec-statmodel}).

\begin{figure*}
	\centering
	\includegraphics[scale=0.75]{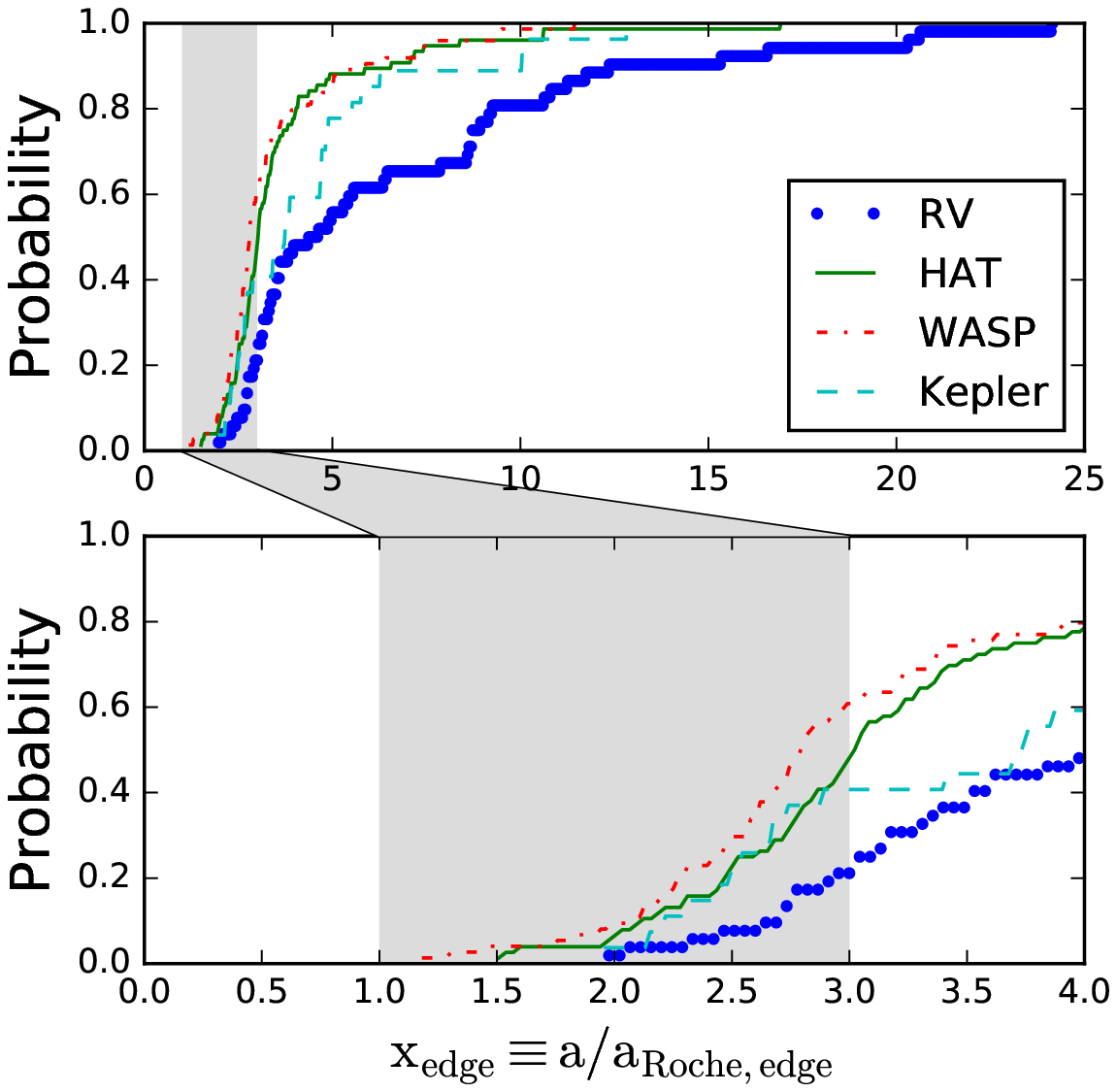}
	\hfill
	\includegraphics[scale=0.75]{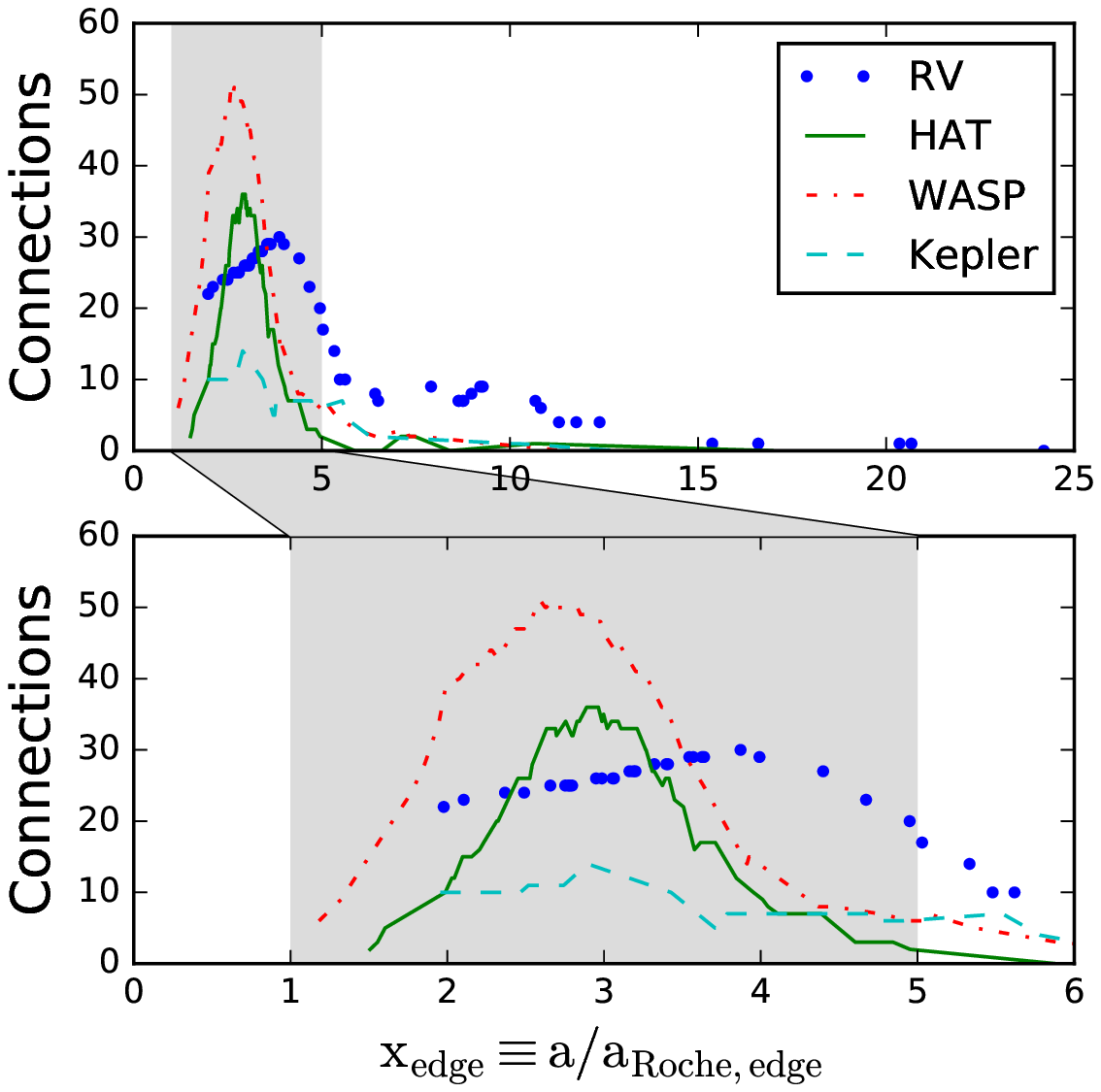}
	\caption{Distribution in $x$ assuming edge-on orbits (i.e., $x_{edge}$) for planets discovered through the following surveys: RV (blue dots), HAT (green solid line), WASP (red dot-dashed line), and Kepler (cyan dashed line). Left: The cumulative distributions across the full range of $x_{edge}$ (top) and zoomed into the $0<x_{edge}<6$ range (bottom). Right: The distribution visualized through a network-based frequency analysis across the full range of $x_{edge}$ (top) and zoomed into the $0<x_{edge}<6$ range (bottom). The effective widths $\zeta$ used here are 1.81 (RV), 0.46 (HAT), 0.84 (WASP), and 0.94 (Kepler). For more on the method, refer to \citet{DerribleAhmad15}. }
	\label{fig-data}
\end{figure*}

We ultimately obtain 52 planets for our RV data, 77 for HAT, 74 for WASP, and 27 for Kepler.
Figure \ref{fig-data} shows how $x$ is distributed for each dataset, assuming edge-on orbits (i.e., the true mass is the same as the measured minimum mass).
At this point, we simply want a rough idea of how these data are distributed, so we visualize them in two different ways: a cumulative distribution to emphasize the location of the smallest $x$ and a more experimental network-based frequency analysis (NFA) to emphasize rough functional form.
The cumulative distribution plots a running probability over $x_{edge}$, i.e., $x$ assuming an edge-on orbit.
The NFA considers each $x_{edge}$ and if they are ``connected'' with other data, rather than counting up $x_{edge}$ in a bin (e.g., as in a traditional histogram), and plots the number of network connections against each value of $x_{edge}$.
Two values are connected if they fall within a given range, $\pm\zeta$, in $x$-space.
The output therefore depends on the chosen value of $\zeta$, but this can be selected in an automated manner by verifying that properties of the network have become stable.
Details of these methods are discussed in \citet{DerribleAhmad15}.\footnote{The code for performing NFA is also publicly available at https://github.com/csunlab/NFA}
The ``optimal'' values of $\zeta$ for these datasets are 1.81 for RV, 0.46 for HAT, 0.84 for WASP, and 0.94 for Kepler.

The cumulative distributions shows $x_{edge}$ truncating around 2 for the RV and Kepler samples, and at $1<x_{edge}<2$ for the HAT and WASP samples.
The NFA agrees with this but also shows the HAT and WASP distributions have a peak somewhere between $2<x_{edge}<3$.
The peak seen in the RV sample may be misleading since most of the data shown do not have physical (but rather minimum) mass measurements.

\begin{figure*}
	\centering
	\includegraphics{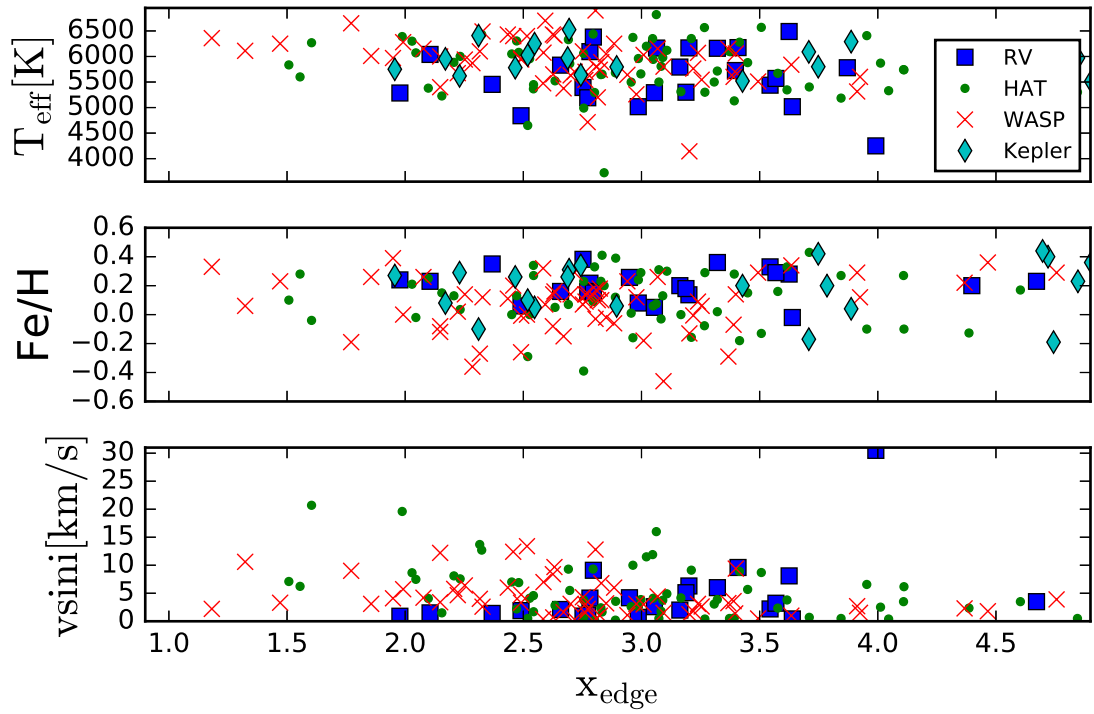}
	\caption{Distribution in $x_{edge}$ vs. various stellar properties (effective temperature [top], metallicity [middle], stellar rotation speed [bottom]) for the RV (blue square), HAT (green dots), WASP (red xs), and Kepler (cyan diamonds) samples.}
	\label{fig-data-stars}
\end{figure*}

Ground-based transit surveys can suffer from selection effects very different from those of RV surveys.
There is a geometric bias toward shorter orbital periods (e.g., the geometric probability of transit scales as $a^{-1}$), transits can be missed due to the rotation of the Earth, three transits are often required to confirm a planet which biases against long-period planets (with respect to the observing baseline), and the stellar demographics may differ from those of RV surveys \citep[and Figure \ref{fig-data-stars}]{Wang15}.

We can easily account for the geometric bias for planets discovered from transit surveys, since $x$ scales with $a$.
The other aforementioned biases depend on details of each survey's observing strategy.
However, the first goal of this study is to not necessarily obtain an accurate statistical description of the true hot Jupiter population(s) but rather to see if the current sample could allow for testing specific models of planet formation.
The observing baseline biases against longer-period planets would mostly affect how the distribution of hot Jupiters tapers off at higher values of $x$.
Fortunately, the HAT and WASP surveys now have sufficient temporal and longitudinal coverage to be sensitive to hot Jupiters over a wide range of orbital periods reaching 30 days.

We can look at a few properties of the host stars sampled from these surveys to verify differences in host star demographics.
Figure \ref{fig-data-stars} looks at correlations in stellar effective temperature, metallicity, and rotation speed with $x_{edge}$ for each survey sample.
We see that RV and Kepler surveys typically probe more metal rich stars than the HAT and WASP surveys.
We also find HAT and WASP planets in the range of $1<x_{edge}<2$ are confined to a narrower region of $T_{eff}$ relative to the overall sample.
We will compare the results of our statistical analysis with these host star properties in \S \ref{sec-discussion}.

\section{Statistical Model}
\label{sec-statmodel}

We approach this problem through a hierarchical Bayesian modeling (HBM) framework.
In HBM, a set of model parameters $\vec{\theta}$ can be explicitly grouped into two subsets, one for parameters describing the individual objects (i.e. individual parameters, $\vec{\alpha}$) and one describing the properties of the objects' population(s) (i.e. population parameters or hyperparameters, $\vec{\beta}$).
Conditioned on the observed data ($\vec{d}$), Bayes theorem is rewritten as
\begin{equation}
p(\vec{\alpha},\vec{\beta}|\vec{d}) = \frac{p(\vec{d}|\vec{\alpha},\vec{\beta}) p(\vec{\alpha}|\vec{\beta}) p(\vec{\beta})}{p(\vec{d})}
\end{equation}
where $p(\vec{d}|\vec{\alpha},\vec{\beta})$ is the likelihood function, $p(\vec{d})$ is the fully marginalized likelihood, and $p(\vec{\alpha},\vec{\beta}|\vec{d})$ is the joint posterior probability distribution.
Using the law of conditional probability, $p(\vec{\alpha}|\vec{\beta}) p(\vec{\beta})$ replaces the prior probability distribution of $p(\theta) = p(\alpha, \beta)$, where $p(\vec{\alpha}|\vec{\beta})$ is the prior on the individual parameters conditioned on the hyperparameters and $p(\vec{\beta})$ is the prior on the hyperparameters.

Explicitly grouping the model parameters in this manner allows us to see the relationship between the individual parameters and hyperparameters.
The distribution of $x$ for each planet is a function of both the individual parameters (e.g. mass and orbital period) and the hyperparameters describing the population.
Likewise, the individual parameters as a group inform the hyperparameters.
This allows for simultaneously inference on the properties of individual planets in our sample and properties of the sample in aggregate.
Recent studies in the exoplanet literature have employed such multi-level modeling to infer the occurrence rate of Earth-analogs \citep{Foreman-Mackey14}, composition distribution of sub-Neptunes \citep{Wolfgang15}, a probabilistic mass-radius relationship for sub-Neptunes \citep{Wolfgang16}, the transition radius between rocky and non-rocky planets \citep{Rogers15}, and the eccentricity distribution of short-period Kepler planets \citep{Shabram16}.

We can obtain posterior distributions on the parameters of interest in this high-dimensional (the length of $\vec{\beta}$ and $\vec{\alpha}$) model through Markov chain Monte Carlo.
We use a Python interface to the statistical library \Stan \citep{Stan16a, Stan16b}, which uses an adaptive Hamiltonian Monte Carlo (HMC) algorithm to sample the posterior distribution.
In essence, HMC samples parameter space by modeling the posterior probability and walker/samplers as a Hamiltonian system.
The posterior probability assumes the role of a potential and the walkers are treated as a particles with some initial kinetic energy.
Steps are proposed using gradient information of the posterior distribution, and the walker is integrated along a path for some finite time.
Although this involves computing a partial derivative in each dimension, the proposal steps are typically much larger than is practical with a random-walk Metropolis-Hastings sampler while retaining a high acceptance probability.
This reduces the effective autocorrelation length of each Markov chain compared to a more traditional random walk Metropolis-Hastings algorithm, resulting in greater efficiency for drawing effectively independent samples.
The efficiency of HMC depends on the chosen integration time.
In fact, a walker may loop back close to its starting position depending on the shape of the posterior distribution, resulting in a small proposal displacement and thus inefficient sampling.
To address this issue, \Stan uses a No-U-Turn Sampler (NUTS) that spawns multiple trajectories and adaptively chooses an appropriate integration length \citep{HoffmannGelman14}.

In the following subsections, we describe two general statistical models that we apply to the data.

\subsection{One-component power law, $\mathcal{M}_1$}

Our first model $\mathcal{M}_{1}$ is inspired by the analysis conducted by \FordRasio.
We model the distribution of $x$ for hot Jupiters as a truncated power law, with power law index $\gamma$ and lower and upper limits of $x$ being $x_l$ and $x_u$ respectively.
This is the same as a Pareto distribution with a lower truncation $x_l$ and upper truncation $x_u$ \citep[``truncated Pareto'' henceforth;][]{Beg81, Aban06}.
Here, $\gamma$, $x_l$, and $x_u$ are the hyperparameters $\vec{\beta}$.

For now, let us assume that all of the observed planets are orbiting edge-on, that all unreported radii are 1.2 $R_J$, and uniform priors on $\gamma$, $x_l$, and $x_u$ over a sufficiently large domain.
Here, the posterior probability distribution on $\gamma$, $x_l$ and $x_u$ conditioned on the observed data $\vec{x}=\{x_1, x_2, \ldots, x_n\}$ of $n$ planets is drawn from (denoted by a $\sim$)
\begin{equation}
p(\gamma, x_l, x_u | \vec{x}) \sim \gamma^n (x_u^\gamma - x_l^\gamma)^{-n} \prod_{j=1}^n x_{j}^{\gamma-1}
\end{equation}
from $x_l$ to $x_u$ and zero elsewhere.
If the planet sample was discovered through a transit survey and we wanted to account for the geometric bias, then we would apply a factor of 1/a (i.e., $\gamma-1$ to $\gamma-2$ in the exponent and $\gamma$ to $\gamma-1$ in the normalization).

For most of radial velocity planets, the orbital inclination and planet radius are not measured.
For convenience, we sort $n$ planets by index $\{1, 2, \ldots, m, \ldots, n-1, n\}$ such that $\{1, 2, \ldots, m\}$ indexes planets with measured inclinations and $\{m+1, \ldots, n-1, n\}$ indexes planets without measured inclinations, with lengths $m$ and $n-m$ respectively.
For these $n-m$ planets, we can model $n-m$ inclinations for the non-transiting planets contained in $\vec{x}$.
For now, unknown radii are still fixed to 1.2 $R_J$.

\begin{figure*}
	\centering
	\includegraphics[scale=0.8]{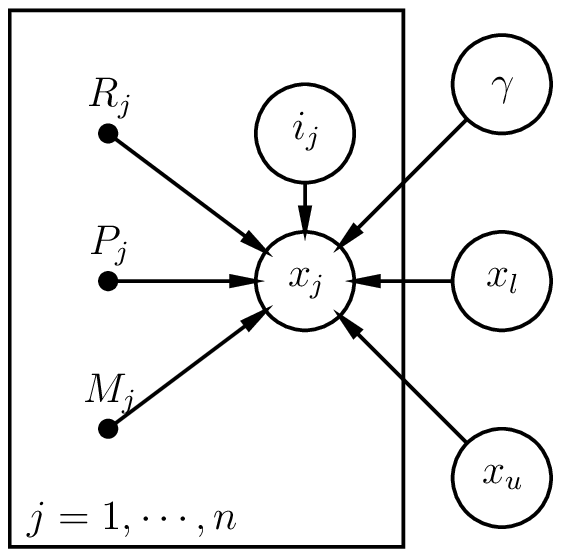}
	\hspace{1.5cm}
	\includegraphics[scale=0.8]{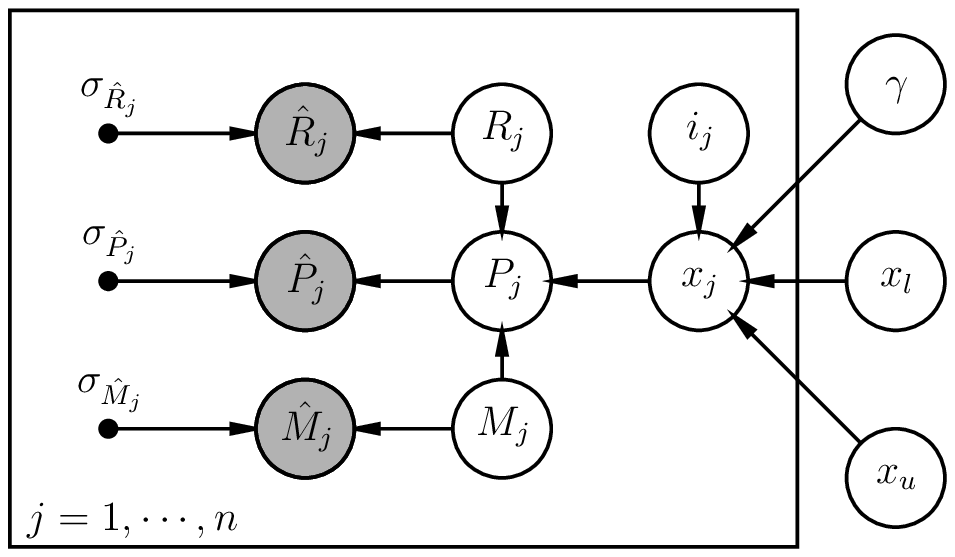} \\
	\vspace{1cm}
	\includegraphics[scale=0.8]{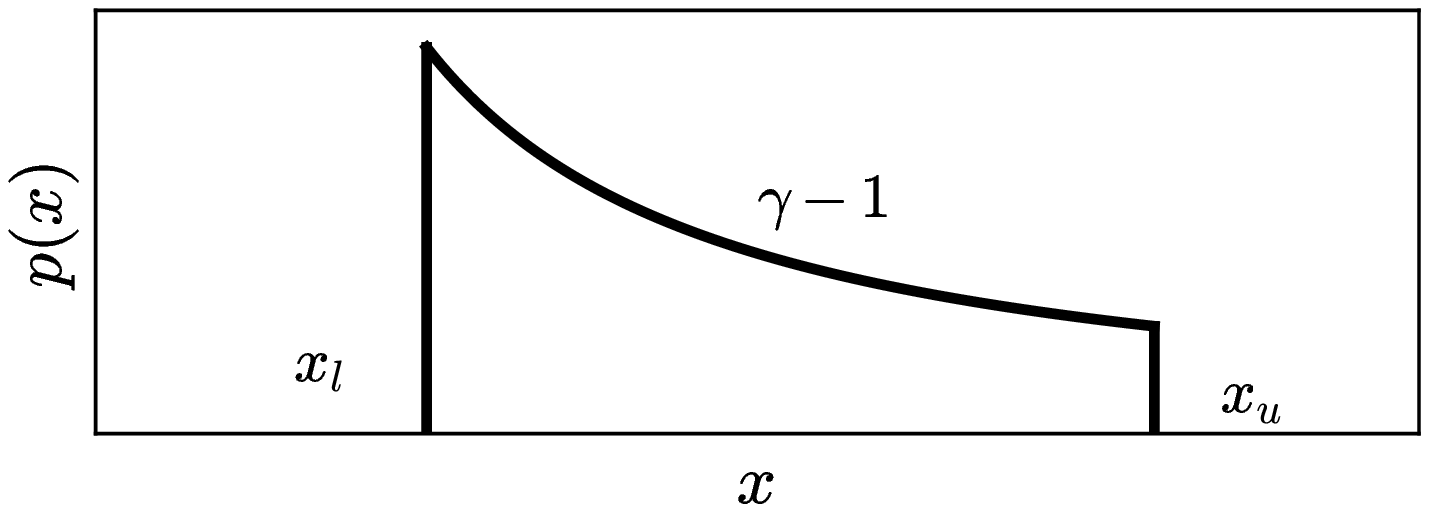} \\
	\caption{
	Probabilistic graphical models for a 1-component truncated Pareto model without ($\mathcal{M}_{1}$) and with ($\mathcal{M}_{1,\sigma}$) observational uncertainties (top).
	Fixed parameters are shown as points, model parameters are open circles, and observations are shaded circles.
	The plate iterates over planets with measured inclinations (indices 1 to $n$).
	Figures was made using $\tt{daft}$ (http://daft-pgm.org).
	Bottom: A cartoon of this model shows the power law index ($\gamma$) and the bounds in $x$ corresponding to the lower ($x_l$) and upper cutoff ($x_u$).
	Refer to Table \ref{tbl-ref} for other variable definitions.}
	\label{fig-pgm1}
\end{figure*}

Our posterior probability distribution is
\begin{equation}
\begin{split}
& p(\gamma, x_l, x_u, i_{m+1}, \ldots, i_n | \vec{x}) \\
& \quad \sim \gamma^n (x_u^\gamma - x_l^\gamma)^{-n}\prod_{j=1}^n \left( \frac{x_{edge,j}}{\sin{i_j}^{1/3}} \right)^{\gamma-1} \\
& \quad = \gamma^n (x_u^\gamma - x_l^\gamma)^{-n}\prod_{j=1}^{m} x_{edge,j} ^{\gamma-1} \prod_{j=m+1}^n \left( \frac{x_{edge,j}}{\sin{i_j}^{1/3}} \right)^{\gamma-1}.
\end{split}
\end{equation}
In this hierarchical framework, \Stan samples from $\vec{\alpha}=\{i_{m+1}, \ldots, i_n\}$ and $\vec{\beta}=\{\gamma, x_l, x_u\}$ simultaneously.
To verify the robustness of this model, we can also incorporate uncertainties in period, mass and radius into a model $\mathcal{M}_{1,\sigma}$.
Here, we assume each observed value ($\hat{P}$, $\hat{M}$, $\hat{R}$) is drawn from a Gaussian distribution with a mean true value ($P$, $M$, $R$) and standard deviation ($\sigma_{\hat{P}}$, $\sigma_{\hat{M}}$, $\sigma_{\hat{R}}$) given by the reported uncertainties.
In other words,
\begin{equation}
\begin{split}
& \hat{P}_j \sim \mathcal{N}(P_j, \sigma_{\hat{P}_j}^2) \\
& \hat{M}_j \sim \mathcal{N}(M_j, \sigma_{\hat{M}_j}^2) \\
& \hat{R}_j \sim \mathcal{N}(R_j, \sigma_{\hat{R}_j}^2)
\end{split}
\end{equation}
where we assume $\sigma_{\hat{R}}=0.1 R_J$ for planets with no radius measurement.
We do not include uncertainties in orbital inclination.
Although posterior distributions obtained from RV or transit data sometimes reveal asymmetric uncertainties in the planetary parameters \citep{Zakamska11}, we focus on period, mass, and radius of planets with high signal-to-noise, so Gaussianity is a reasonable assumption for this study.
This appears to be the case for the vast majority of reported uncertainties in masses and orbital period.
We do note a varying degree of asymmetry in some of the radius measurement uncertainties.
In those cases, we simply average the upper and lower error bar as our effective $\sigma_{\hat{R}}$.
Figure \ref{fig-pgm1} shows $\mathcal{M}_{1}$ and $\mathcal{M}_{1,\sigma}$ in a probabilistic graphical form (neglecting differences in $i$ and $R$ for planets $1,\ldots,m$ and $m+1,\ldots,n$ for simplicity).

A truncated Pareto distribution is convenient here for a few reasons.
First, it provides a direct comparison to the results of \FordRasio, who employed the same distribution.
Second, it is unclear how nature distributes hot Jupiters in $x$-space.
A truncated Pareto is flexible enough to allow for monotonically increasing and decreasing (or uniform) frequency in $x$ over the limited range of our sample.
Lastly, the lower truncation parameter $x_l$ is motivated as a well-understood theoretical stopping point for planets undergoing disk or eccentric migration at $x_l=1$ and $x_l=2$ respectively.
However, we discuss some caveats in \S\ref{sec-discussion}.

\subsection{Two-component power law mixture}

It is not clear that a single-component Pareto would accurately describe our datasets based on Figure \ref{fig-data}, specifically the HAT and WASP sample.
Furthermore, there may be enough planets in aggregate to meaningfully constrain multiple overlapping populations.
In particular, what if the current sample of hot Jupiters were the result of a mixture of two processes: one component consistent with disk migration and the other with eccentric migration?

\begin{figure*}
	\centering
	\includegraphics[scale=0.8]{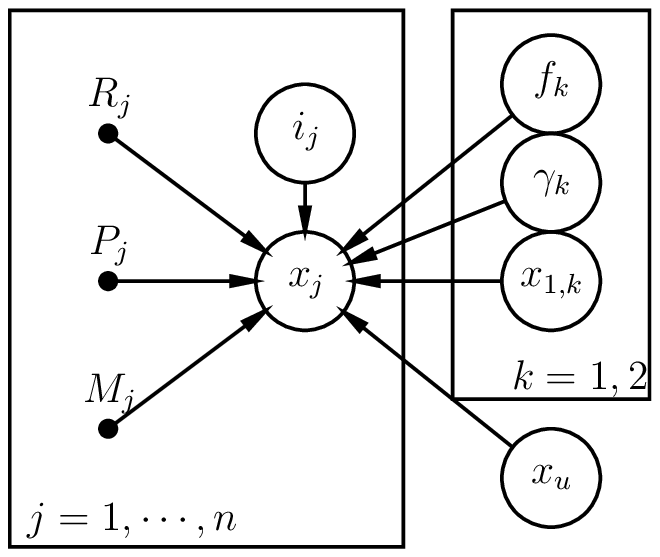}
	\hspace{1.5cm}
	\includegraphics[scale=0.8]{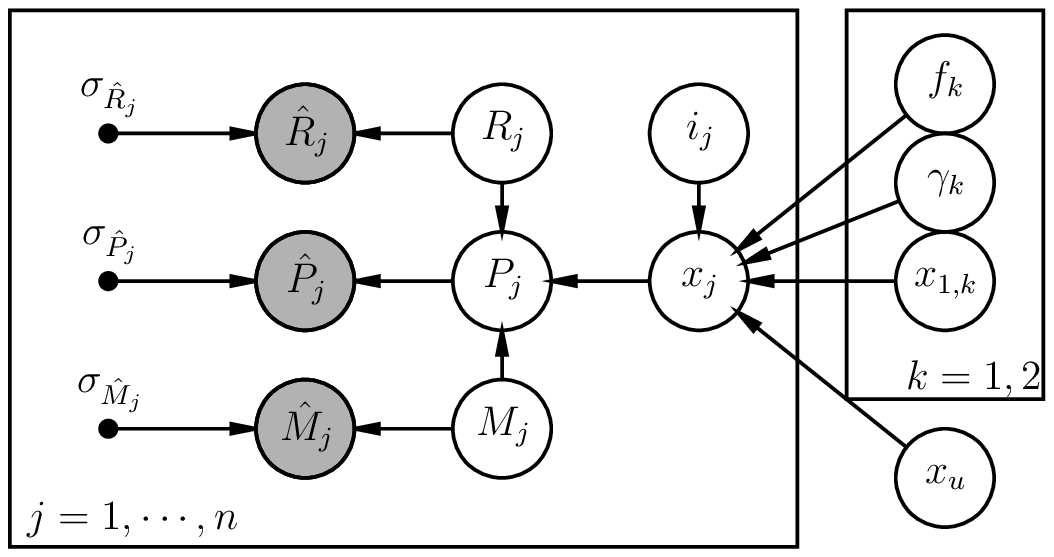} \\
	\vspace{1cm}
	\includegraphics[scale=0.8]{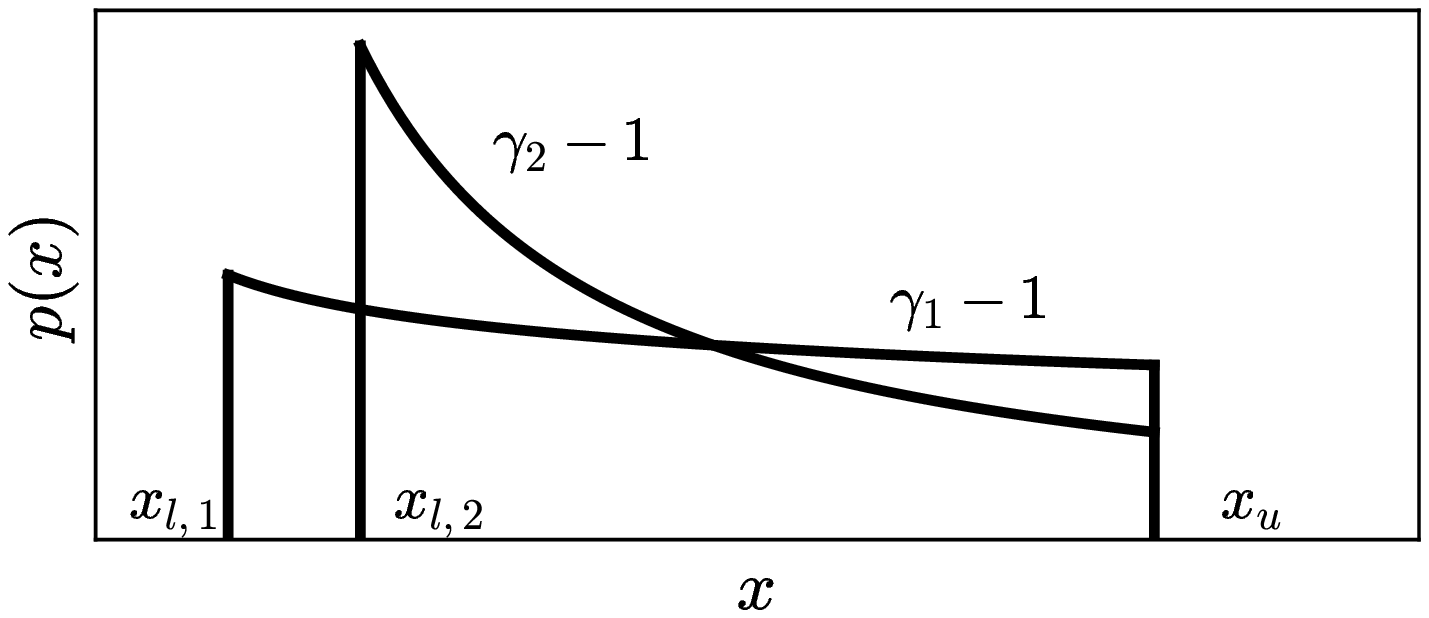} \\
	\caption{
	Probabilistic graphical models for a 2-component truncated Pareto model without ($\mathcal{M}_{2}$) and with ($\mathcal{M}_{2,\sigma}$) observational uncertainties (top).		Figures was made using $\tt{daft}$ (http://daft-pgm.org).
	Bottom: A cartoon of this model which is similar to Figure \ref{fig-pgm1} but now has two power-law indices ($\gamma_1$ and $\gamma_2$) and an additional $x_l$ parameter.
	Refer to Table \ref{tbl-ref} for variable definitions.}
	\label{fig-pgm2}
\end{figure*}

Here, we consider a model with a mixture of two truncated Pareto distributions, $\mathcal{M}_{2}$.
Each component $k$ is modeled with its own $\gamma_k$, $x_{l,k}$, and the fraction of planets residing in the $k$th component $f_k$.
$f$ is a simplex demanding that $f_1+f_2=1$.
Both components share a common $x_u$.
We therefore sample from the following posterior distribution:
\begin{equation}
\begin{split}
& p(\vec{f}, \vec{\gamma}, x_{l,1}, x_{l,2}, x_u, i_{m+1}, \ldots, i_n | \vec{x}) \\
& \quad \sim \prod_{j=1}^n \Biggl\{ f_1 \frac{\gamma_1}{(x_u^{\gamma_1} - x_{l,1}^{\gamma_1})} \left( \frac{x_{edge,j}}{\sin{i_j}^{1/3}} \right)^{\gamma_1-1} \\
& \quad + f_2 \frac{\gamma_2}{(x_u^{\gamma_2} - x_{l,2}^{\gamma_2})}\left( \frac{x_{edge,j}}{\sin{i_j}^{1/3}} \right)^{\gamma_2-1} \Biggr\}.
\end{split}
\end{equation}
Figure \ref{fig-pgm2} shows the probabilistic graphical models and visualizes the hyper parameters for $\mathcal{M}_{2}$ and $\mathcal{M}_{2,\sigma}$.

\section{Results}
\label{sec-results}
We apply $\mathcal{M}_{1}$ and $\mathcal{M}_{2}$ to the RV+Kepler data and HAT+WASP data separately and sample from the posterior distribution using \Stan.

We assess the evidence for both models by computing the fully marginalized likelihood (i.e., Bayesian evidence) of the population-level parameters using an importance sampling algorithm \citep{Benelson16}.
For the RV+Kepler data, the Bayes factor $p(\vec{d}| \mathcal{M}_{2})/p(\vec{d}| \mathcal{M}_{1})$ is $\approx$0.1.
For the HAT+WASP data, the Bayes factor is on the order of $10^{21}$.
Given the choice of our statistical models, we conclude that the RV+Kepler data do not provide sufficient evidence for multiple populations, but the HAT+WASP data show decisive evidence for two populations.

For the RV+Kepler data, we run our MCMC with 10 chains for at least 300,000 generations each for $\mathcal{M}_{1}$ and $\mathcal{M}_{1,\sigma}$.
For the HAT+WASP data, we run our MCMC with 10 chains for at least 400,000 generations each $\mathcal{M}_{2}$ and $\mathcal{M}_{2,\sigma}$.
We throw out the first half of the sample as burn-in then thin the remainder to ultimately obtain 10,000 posterior samples.

\begin{figure*}
	\centering
	\includegraphics{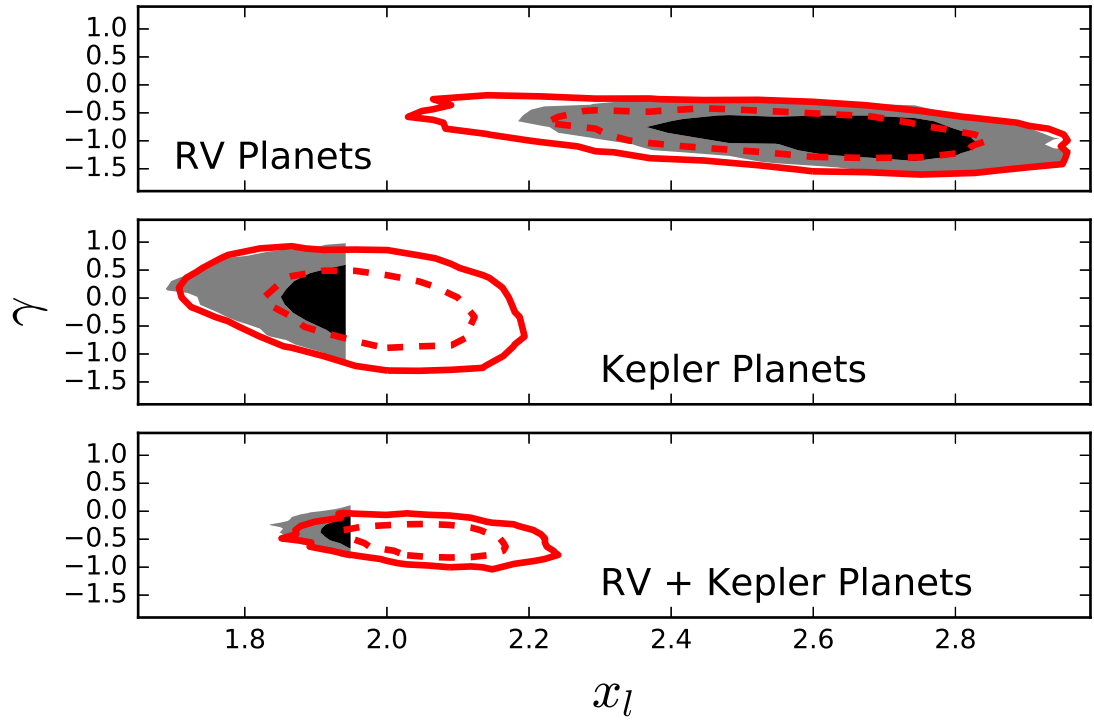}
	\caption{Joint posterior distributions of the power law index $\gamma$ and $x_l$ for a one-component model (see Figure \ref{fig-pgm1}) for a dataset containing solely RV-discovered planets (top panel), Kepler planets (middle panel), and RV+Kepler planets (bottom panel). Filled contours show 68\% (black) and 95\% (gray) credible intervals without considering uncertainties in observed planet properties (e.g. radius, orbital period, mass). Line contours show 68\% (dashed) and 95\% (solid) credible intervals with uncertainties in observed planet properties.}
	\label{fig-results-rvkepler}
\end{figure*}

\begin{figure*}
	\centering
	\includegraphics[scale=0.6]{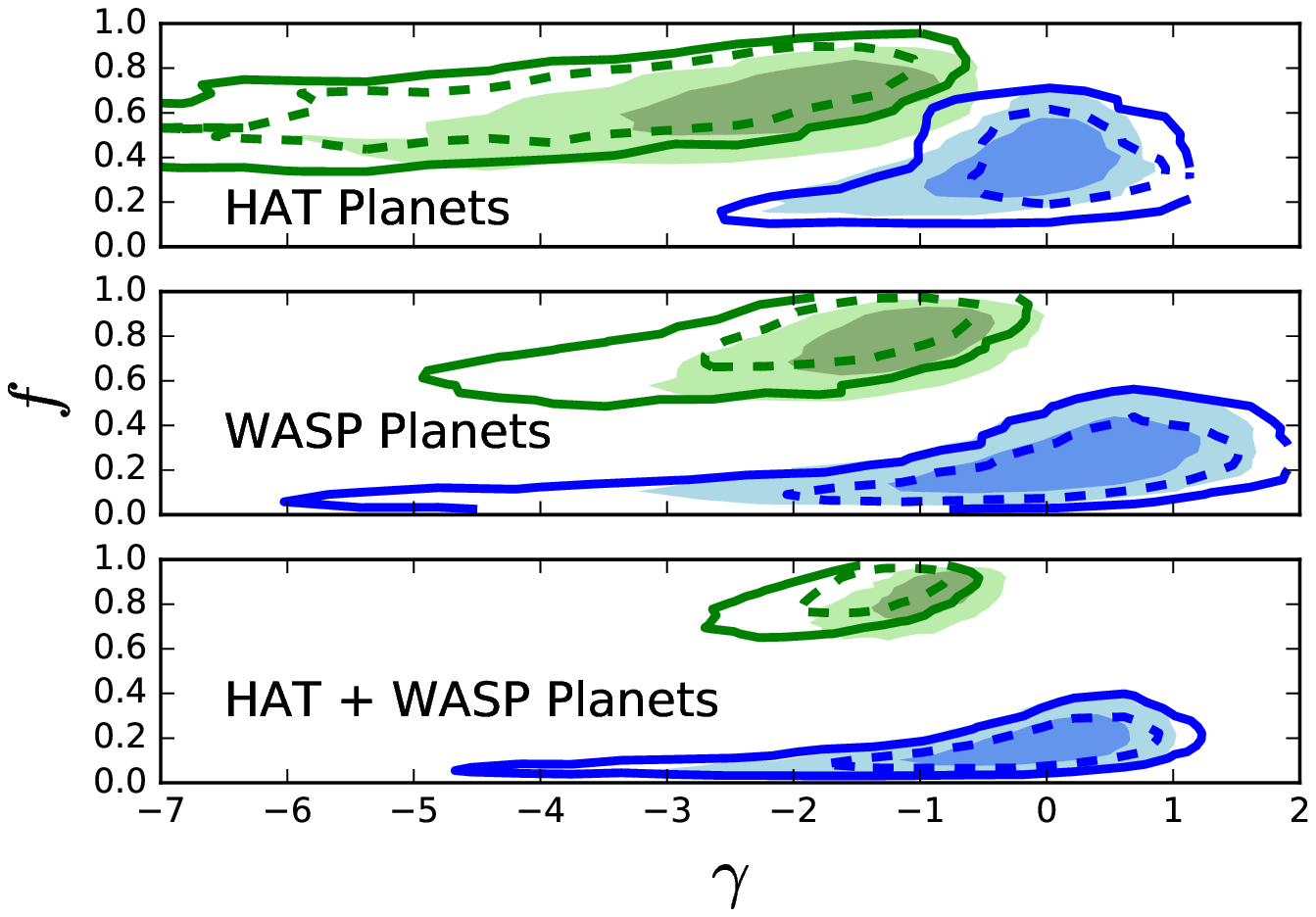}
	\hspace{1cm}
	\includegraphics[scale=0.6]{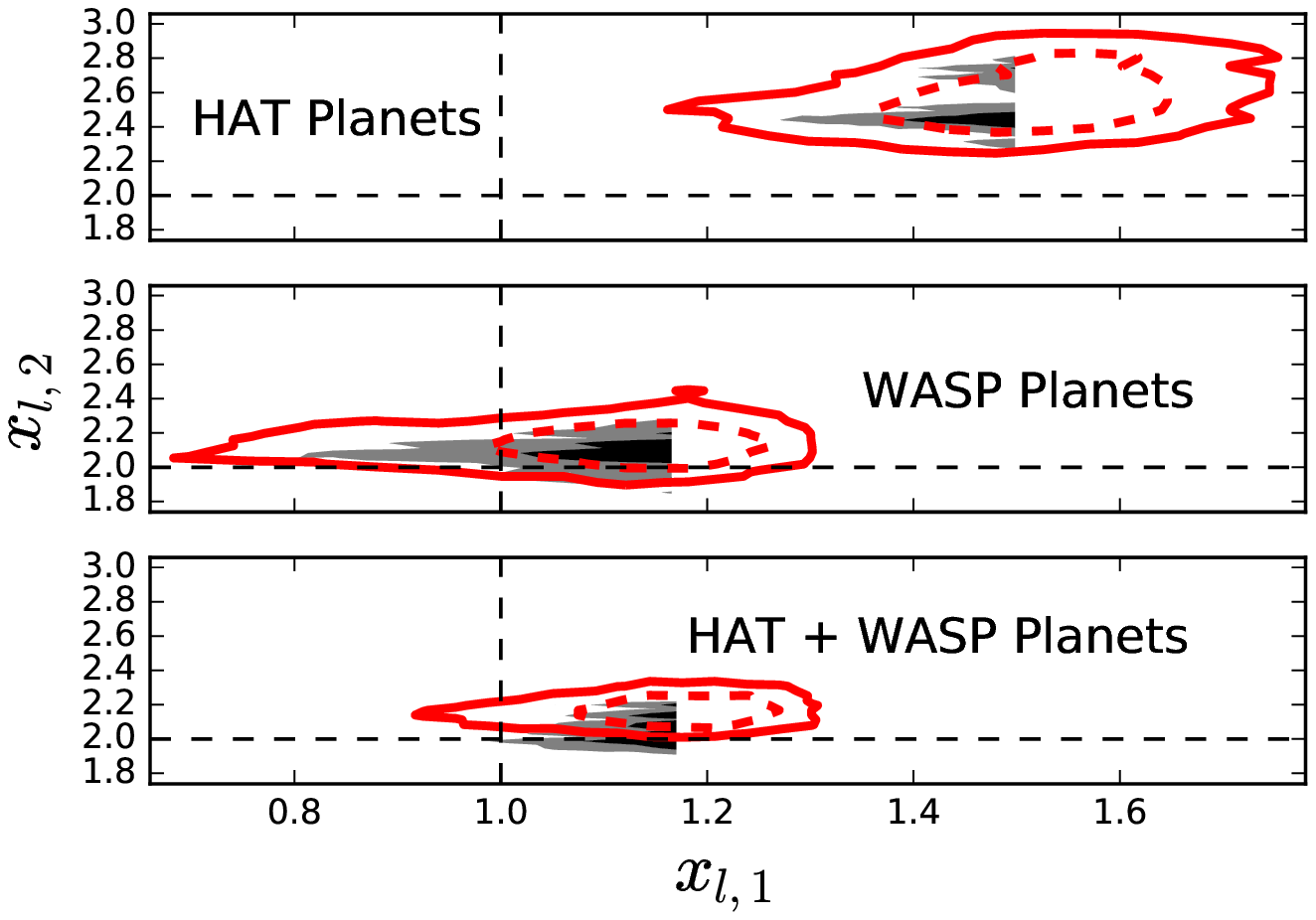}
	\caption{ Joint posterior distributions of a two-component model (see Figure \ref{fig-pgm2}) for dataset containing solely WASP planets (top panels) and HAT+WASP planets (bottom panels). Both figures show filled contours show 68\% (darker) and 95\% (lighter) credible intervals without considering uncertainties in observed planet properties. Line contours show 68\% (dashed) and 95\% (solid) credible intervals with uncertainties in observed planet properties. \textbf{Left}: Joint posterior distributions of power law index $\gamma$ (horizontal axis) and component fraction $f$ (vertical axis) for the first (e.g. $x_{l,1}$) and second (e.g. $x_{l,2}$) components in blue and green respectively. \textbf{Right:} Joint posterior distribution of $x_l$ for the first (horizontal axis) and second (vertical axis) components compared to the theoretical truncation locations of the two migration models (dashed lines).}
	\label{fig-results-hatwasp}
\end{figure*}

For Figures \ref{fig-results-rvkepler} and \ref{fig-results-hatwasp}, filled contours correspond to models that do not include uncertainties in measurements of the planet properties (i.e, $\mathcal{M}_{1}$ or $\mathcal{M}_{2}$) and line contours correspond to models that do include observational uncertainties (i.e., $\mathcal{M}_{1,\sigma}$ and $\mathcal{M}_{2,\sigma}$).

\subsection{One-component model: RV and Kepler datasets}

The functional form of $p(\beta|d)$ for the RV and Kepler data look similar, ending abruptly at $x\approx2$ and generally tapering off at larger $x$.
Figure \ref{fig-results-rvkepler} shows the posterior distribution of $\gamma$ and $x_l$ for the one-component model applied to the these datasets individually and together.

In the RV sample, there are a couple planets with $x_{edge}\approx2$, but this is assuming their minimum masses are close to their physical masses.
By modeling unknown inclinations (thus unknown physical masses) for these planets and all the planets as drawn from a single population, we find a probable range of $x_l$ from 2.1 to 2.9.
In the Kepler sample, all the planets are transiting, so their RVs probe physical masses.
$x_l$ is governed by the planet with the smallest $x$, which is slightly less than 2, and we see a truncated contour.
In the RV+Kepler sample, the location of $x_l$ remains the same as the Kepler-only result but we obtain a more precise constraint on $\gamma$.
We can verify the robustness of these results by modeling measurement uncertainties, and they indeed overlap very well with $\gamma=-0.51\pm^{0.19}_{0.20}$.
In each case, there is an anti-correlation between $x_l$ and $\gamma$, which we interpret as an artifact of our chosen population model: tuning $x_l$ to higher values for a fixed $x_u$ forces the distribution to become much steeper, i.e., a $\gamma$ that is more negative.

The above result appears to be consistent with a single population that underwent an eccentric migration history.
However, we note that the RV sample includes several hot Jupiters in multi-planet systems.
Some even have multiple planets with semi-major axes less than 1 AU (e.g. 55 Cancri, HIP 14810).
We perform a separate run on an RV dataset without known multi-planet systems and find quantitatively consistent results in regards to $x_l$ and $\gamma$.

We additionally test $\mathcal{M}_{2}$ for the combined dataset and find the population-level parameters to be mostly unconstrained.
This is reflected in our Bayes factor which shows a lack of strong evidence for $\mathcal{M}_{2}$.

After a decade since \FordRasio, it appears the inner edge at $x_l\approx2$ still holds for RV discovered planets, and maybe even Kepler planets.
We will address possible caveats regarding this in \S \ref{sec-discussion}.

\subsection{Two-component mixture model: HAT/WASP and all datasets}
\label{sec-results-2comp}

Figure \ref{fig-results-hatwasp} shows the posterior distribution of several parameters that specify our two-component mixture model applied to the HAT+WASP data.
Our MCMC was initialized around $x_{l,1}\sim1$ and $x_{l,2}\sim2$, motivated from the predictions made by disk and eccentric migration respectively.
Two-component properties of the HAT and WASP samples could be meaningfully constrained, as shown in the top and middle panels of both figures.
The results for the HAT+WASP samples are shown in the bottom panels.

The left figure shows constraints on the component fractions $f$ and power law indices $\gamma$ of each modeled component (blue and green contours).
Neither component has $f$ consistent with 0 or 1, suggesting our model can successfully separate two overlapping populations of planets.
For the model including uncertainties applied to the HAT+WASP data, estimates for component fractions based on 15.9, 50, and 84.1 percentiles are $0.15\pm^{0.09}_{0.06}$ and $0.85\pm^{0.06}_{0.09}$ for the first and second components respectively.
The first component suggests a relatively shallow Pareto distribution over $x$ ($\gamma=-0.04\pm^{0.53}_{1.27}$) and the second component suggests a much steeper distribution toward smaller $x$ ($\gamma_2=-1.38\pm^{0.32}_{0.47}$).

The right figure shows constraints on each $x_l$ compared to the theoretical truncation locations of the two migration models (dashed lines).
The posterior distribution of $x_{l,1}$ is unsurprisingly governed by the planet with the smallest $x$, but $x_{l,2}$ shows multi-modal structure around $x_{l,2}\approx2$ (filled contours).
The distribution looks smoother once observational uncertainties are considered (line contours).
Some Markov chains were found to be sampling more confined regions of $x_{l,2}$ space and suspect these were local minima.
We simply removed these chains from our posterior sample.

We can also quantitatively assess convergence of the hyperparameters using the Gelman-Rubin statistic.
In practice, a value close to 1 implies the estimand has converged.
We compute the following values for each hyperparameter: 1.035 ($f$), 1.046 ($\gamma_1$), 1.038 ($\gamma_2$), 1.004 ($x_{l,1}$), 1.085 ($x_{l,2}$), and 1.020 ($x_u$).
We provide our interpretations in \S \ref{discussion-results}.

\subsection{Two-component model: synthetic dataset}
\label{sec-results-synth}
Could the multi-modality issues in the previous section be attributed to a relatively small or incomplete sample of hot Jupiters?
To test this, we attempt to recover the input properties of three synthetically generated datasets using a two-component mixture model (without uncertainties).
These datasets contain a set of planets with known inclinations drawn from a mixture of two truncated Paretos with the following properties: $\{f_1, f_2\} = 0.25, 0.75$; $\{\gamma_1, \gamma_2\} = 1, -1$; $\{ x_{l,1}, x_{l,2}\} = 1, 2$; and $x_u = 20$.
The sample size is 100, 500, and 1000 respectively, and our MCMC is initialized close to the true values of the generative model.

\begin{figure*}
	\centering
	\includegraphics[scale=0.6]{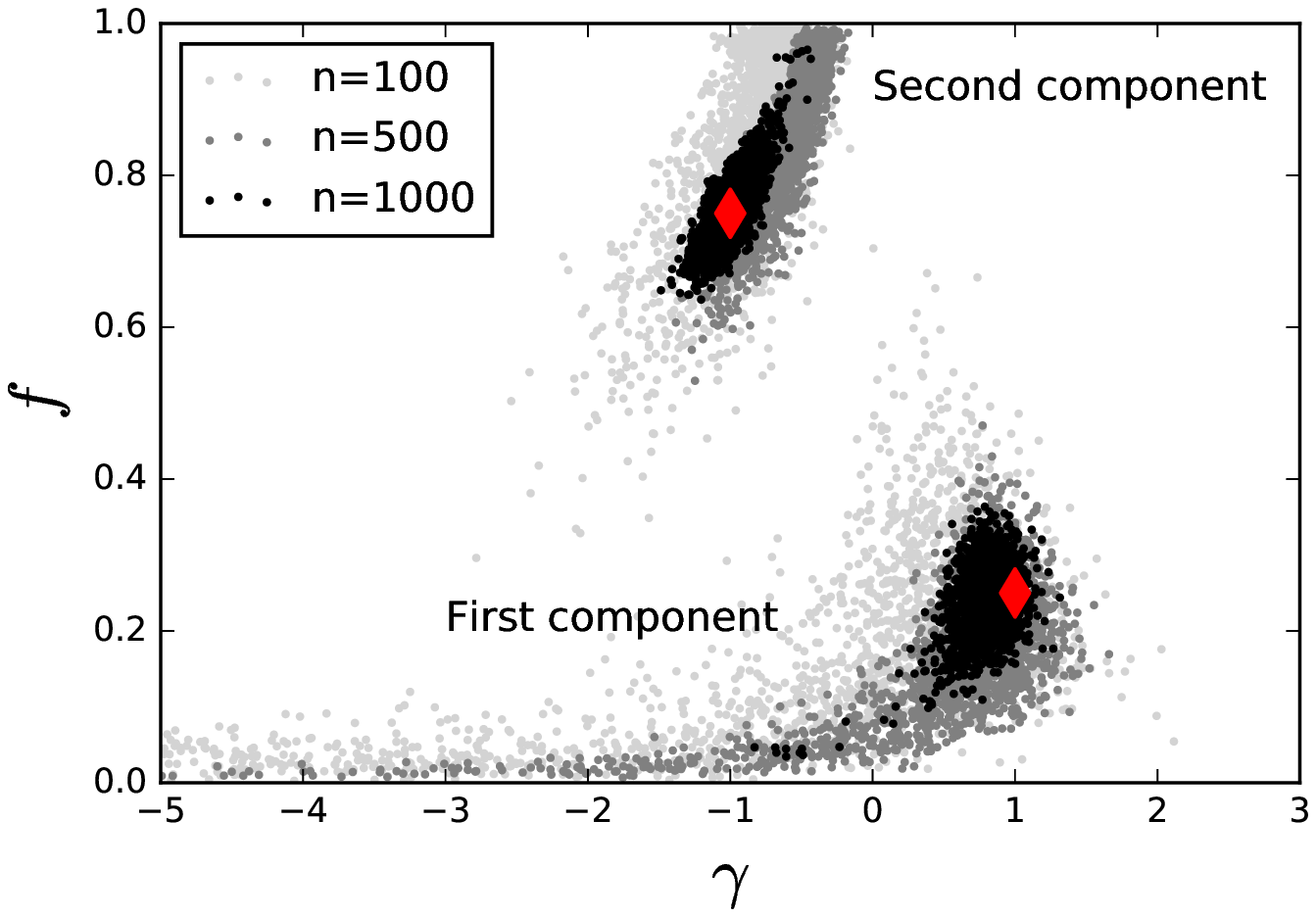}
	\hspace{1cm}
	\includegraphics[scale=0.6]{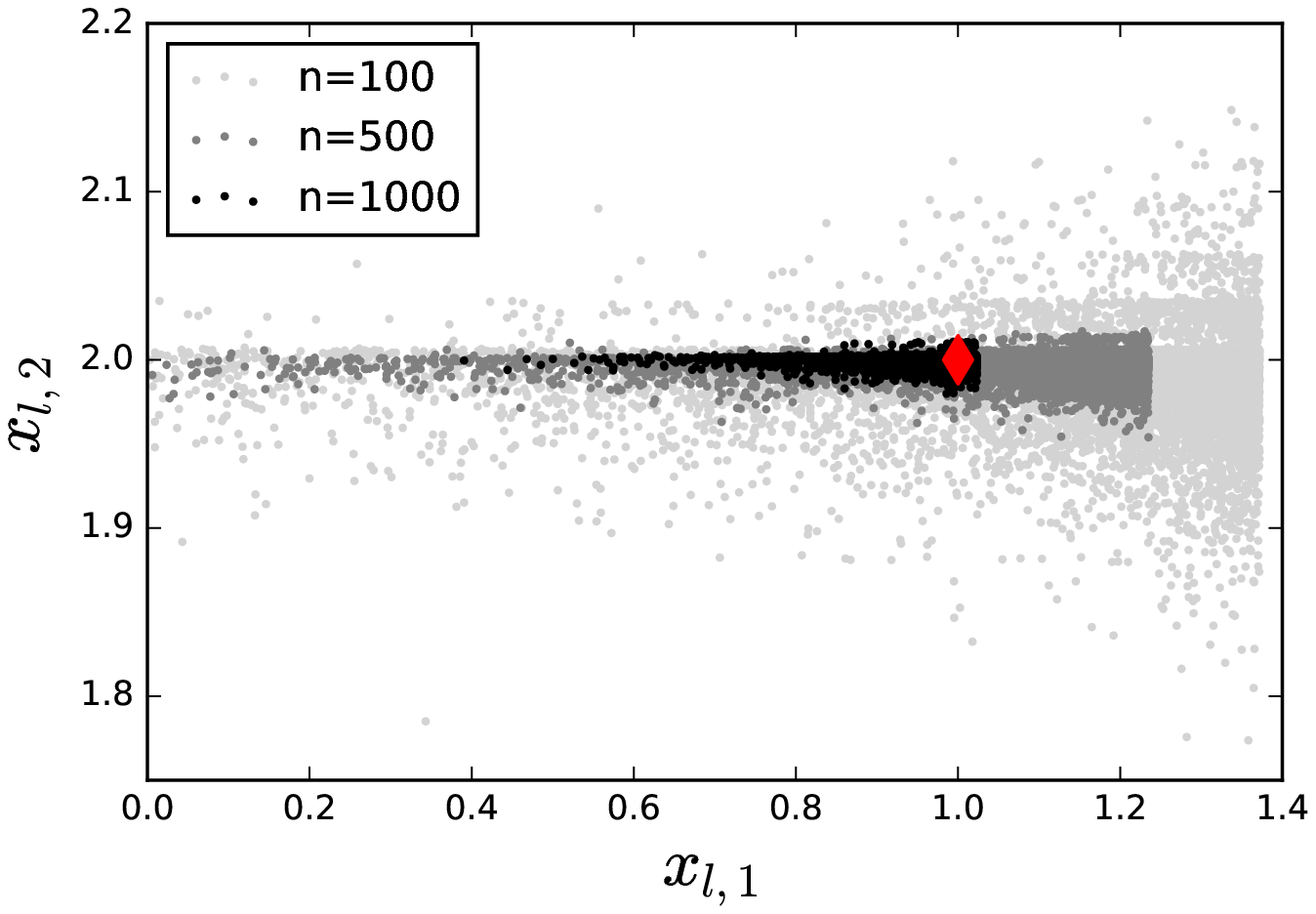}
	\caption{Joint posterior distributions of a two-component model (see Figure \ref{fig-pgm2}) for a synthetic dataset containing a mixture of populations (see \S \ref{sec-results-synth}). Both figures show scatterplots of population parameters when considering a dataset of 100 (light gray), 500 (gray), and 1000 (black) planets. The true, input properties of the generative model are denoted by the red diamonds. \textbf{Left}: Joint posterior distributions of power law index $\gamma$ (horizontal axis) and component fraction $f$ (vertical axis) for the first (near bottom cluster) and second (near top cluster) components. \textbf{Right:} Joint posterior distribution of $x_l$ for the first (horizontal axis) and second (vertical axis) components.}
	\label{fig-results-synth}
\end{figure*}

Figure \ref{fig-results-synth} show constraints on the same hyper-parameters discussed in \S \ref{sec-results-2comp} for our three datasets in comparison to the true input values denoted by red diamonds.
In each case, our constraints were consistent with the true values of the generative model.
As was shown in previous figures, the distribution of $x_{l,1}$ is truncated at the value of the smallest $x$ in each dataset.
Some multi-modal structure arises in $x_{l,2}$ for the dataset with only 100 planets, somewhat similar to what was seen in Figures \ref{fig-results-hatwasp}.
This structure disappears when the sample size increases.
So it is plausible that the multi-modal structure seen in the posterior for the real data can be at least partially attributed to sample size.

\section{Discussion}
\label{sec-discussion}
The first conclusion we draw from this study is that the multiple populations can be inferred from the current sample of hot Jupiters given the limitations of our hierarchical Bayesian framework.
This suggests that one could potentially disentangle populations consistent with having a disk or eccentric migration history, perhaps placing informative constraints on the prevalence of each mechanism given a large enough sample size (at least a few hundred).
Of course, there are many caveats to address in regards to the datasets, model assumptions, and interpretations of the final results.

\subsection{The Data}
\label{discussion-data}

\subsubsection{Choosing the NASA Exoplanet Archive vs. \exoplanetsorg}

We present results based on planets listed in the NASA Exoplanet Archive but a similar analysis applied to another exoplanet database provides an important test on how sensitive our results are to the criteria for which planets are included in these datasets.

There are a few differences in the planets queried from the NASA Exoplanet Archive and \exoplanetsorg when using similar filters.
The most substantial difference was in the number of HAT planets: 51 for \exoplanetsorg and 77 for the Exoplanet Archive.
The former contains many more HATSouth planets than the former, but the cumulative $x$ distributions are very similar.
The Exoplanet Archive also finds fewer WASP planets than \exoplanetsorg.
Some WASP planets do not have a listed RV semi-amplitude, and these are missed when we filter for $K>30$\ms.
Other exoplanets including HD 189733 are also missed by the Exoplanet Archive due to this semi-amplitude filtering issue.


\subsubsection{The NASA Exoplanet Archive dataset}

Focusing on the RV planets specifically, we find that $x_l\approx2$ holds even after a decade of new discoveries.
The Kepler data also seem to show a similar cutoff, albeit with a smaller sample.

On one hand, this may mean that this subset of Kepler stars and those probed by RV are similar or at least share common conditions of their local galactic environments.
On the other hand, the results of our analysis of the HAT+WASP data may suggest that we are missing some very short-period hot Jupiters in the RV or Kepler data.
For the latter, it is known that part of the Transit Planet Search software module searches for and removes a set of known harmonics in the flux time series \citep{Jenkins10b, Tenenbaum12}.
Some very short-period transits may resemble these harmonics and be accidentally removed from these data.
If this has been the case, there could be a handful of $1<x<2$ hot Jupiters around relatively bright stars awaiting discovery in the Kepler data.
Another possibility for the lack of $1<x<2$ planets in the RV and Kepler data may stem from insufficient sampling of the true population.

The HAT+WASP data show an increase in planet frequency until $x\approx3$ then a tapering at higher values.
Transit surveys should be mostly biased against detecting planets with larger $x$, so the physical meaning behind the peak around $x\approx3$ is unclear.
If such biases in these long-running surveys are not negligible, we would expect that the mode of this hot Jupiter population shifts to slightly higher values and the distribution would have a stronger tail toward larger $x$ values.

\subsection{The Model}
\label{discussion-model}

Simple theories of planet formation generally fail to explain the wide diversity of planetary systems seen to date.
Many of the model assumptions were motivated more by statistical convenience rather than an accurate reflection the observed hot Jupiter population(s).

For instance, many of the radius measurements had very asymmetrical errorbars, ranging from small to order of magnitude differences. 
For simplicity, we modeled the observed radii as being drawn from a symmetric Gaussian distribution.
Future work can improve upon this using a more general prescription of modeling asymmetric uncertainties (e.g., using a skewed Gaussian) or using the joint posterior distributions of $P$, $M$, and $R$ if available.

For each dataset, we set unknown radii to 1.2 $R_J$ and $\sigma_{\hat{R}}=0.1 R_J$.
\FordRasio found a near perfect degeneracy between $R$ and $x_l$ in the RV sample, even when including known transiting planets, so we expect this relation to hold if we chose a different $R$.
In other words, setting unknown radii to a larger (smaller) mean value would simply decrease (increase) $x_l$ by a proportional amount.

A major limitation of the truncated Pareto is that $x_l$ largely depends on the planet with the smallest $x$.
Imposing such a hard edge may not offer enough flexibility with the detailed physics suggested by the observations.
Eccentric migrating planets that circularize at $x\approx2$ may experience further orbital decay due to long-term tidal dissipation within the slowly-rotating host star \citep{ValsecchiRasio14b}.
This orbital decay timescale can be very sensitive to the planet's semi-major axis depending on the tidal evolution prescription \citep[e.g, $a^8$,][]{ValsecchiRasio14a}.
This would smear out the lower truncation and create a low probability tail inward of x=2 that these surveys may not have probed yet.
So for example, if one were to motivate that the observations were consistent with a population with an eroded edge, they could instead model a power law with an inner ``soft'' edge, perhaps a piecewise function parameterized with a linear slope or exponential decay.
This could provide an alternative explanation for the observations and should certainly be investigated in a future study.

Our primary interest here is to look at these data in the context of the two well-studied hot Jupiter formation channels and lay the groundwork for more complex modeling of single or overlapping populations.
Any detailed hot Jupiter formation theory needs to motivate a particular population-level distribution to sample from, and such an exhaustive investigation of these theories and how to model their representative distributions is beyond the scope of this paper.

\subsection{Interpreting The Results}
\label{discussion-results}
Sampling from the single-component model was computationally tractable and the Markov chains appear to be well sampled.
The two-component model proved to be a much greater computational challenge.
At first glance of the parameters in Figure \ref{fig-results-hatwasp}, it seems the MCMC suffered from sampling issues in the model that neglected uncertainties, since multiple modes were found in $x_{l,2}$ space.
We find these modes have comparable posterior probabilities, so the data do not seem to favor a unique unimodal solution.
Nevertheless, \Stan was able to sample from multiple modes quite efficiently, and the Gelman-Rubin statistic shows multiple chains were able to converge to similar distributions.

Other than the location of the inner cutoffs, we also place constraints on $\gamma$, $\gamma_1$, and $\gamma_2$.
For the Kepler and RV samples, we find $\gamma=-0.51\pm^{0.19}_{0.20}$ for a single-component Pareto.
We compare this to $\gamma_2=-1.38\pm^{0.32}_{0.47}$ for the second component of two-component mixture model applied to the HAT+WASP samples.
These estimates are consistent to within 2-sigma, but this can be interpreted in a few ways.
First, there is an anti-correlation between $x_l$ and $\gamma$, since increasing $x_l$ for a fixed $x_u$ will push $\gamma$ to more negative values.
This is most easily seen in the top panel of Figure \ref{fig-results-rvkepler} for the RV data alone, where our estimate of $x_l$ overlaps much better with that of $x_{l,2}$.
In this case, the power law indices are consistent to within 1-sigma.
Second, perhaps non-negligible observing biases exist in the HAT+WASP samples that negatively affect the detection of planets at larger $x$ and accounting for these would likely increase $\gamma_2$, matching up with the RV+Kepler results much better.
In any case, the uncertainty in $\gamma_2$ is moderately large, and it is unclear if the distributions associated with $\gamma$ and $\gamma_2$ probe similar populations of planetary systems.

After applying our statistical models to the \exoplanetsorg data, we arrive at the similar conclusions regarding our population inference.
The RV+Kepler data do not find significant evidence for $\mathcal{M}_{2}$ (Bayes factor $\approx 0.1$).
For $\mathcal{M}_{1}$, we find $x_l \approx 2$ and $\gamma=-0.53\pm0.23$, which is consistent with the Exoplanet Archive planets.
The HAT+WASP data find decisive evidence for $\mathcal{M}_{2}$ (Bayes factor $\approx 10^{14}$) with moderately different $\gamma$ values and mixture fractions.
We find that $35\pm10\%$ reside in the disk migration component ($\gamma=0.82\pm^{0.28}_{0.34}$) and $65\pm10\%$ are in the high-eccentricity migration component ($\gamma=-2.45\pm^{0.94}_{1.34}$).

Given the evidence for two populations, it is natural to ask whether the mixture fractions correlate with other observable properties, such as the host star temperature, metallicity or rotation rate.
Based on an exploratory analysis of the Exoplanet Archive data set, we identify a potential difference in the $x$ distribution of planets with host star effective temperature.
Planets with $1<x<2$ seem to be preferentially around hotter stars.
We caution that there is a relatively small number of planets in this range and that this apparent difference is not replicated when we perform a similar analysis of the \exoplanetsorg dataset.
This demonstrates that, despite enormous progress in these exoplanet surveys, there is still motivation for further increasing the sample size in order to address basic questions about the formation of hot Jupiters.

\subsection{Looking forward}
Upcoming exoplanet missions will significantly expand the hot Jupiter sample.

The KELT survey \citep{Pepper07, Pepper12} has published over a dozen hot Jupiters, with at least one near tidal disruption \citep{Oberst16}, and is increasing sensitivity to longer-period planets.
The Transiting Exoplanet Survey Satellite \citep[TESS,][]{Ricker15} is an all-sky mission designed to target nearby bright stars that are amenable to ground-based transit and RV follow-up.
TESS will stare at 13 ``sectors'' over the course of a year, which should yield on the order of 70 hot Jupiters around pre-selected stars observed with a 2-minute time sampling and on the order of 10,000 around stars observed in full-frame images with a 30-minute sampling \citep{Sullivan15}.
However, this survey design means that stars closer to the ecliptic poles will receive a longer observational baseline, enabling longer period transits to be characterized.
This observational bias may need to be accounted for when studying hot Jupiters from TESS, depending on the assumed orbital period cutoff that defines this population.
Furthermore, getting mass measurements of every TESS hot Jupiter would be an intractable observational campaign.
In principle, one could invoke a probabilistic mass-radius relation \citep[e.g.,][]{Wolfgang16,Chen16}. 
Since $R$ depends very weakly on $M$ in the known giant planet regime, this could be offset with a large enough sample size and $x$'s weak dependence on $M$.

Looking into the farther future, \emph{WFIRST} will measure precise light curves for millions of stars with the intent of finding long-period or free-floating microlensing planets.
However, this survey also has an expected yield of several thousand transiting hot Jupiters \citep{Montet16}.
These stars would most likely be too faint to make precise planetary mass measurements.
However, the population statistics of these hot Jupiters residing near the Galactic Center can be directly compared to studies of those in the field, which can reveal details on hot Jupiter formation as a function of local galactic environment.

All of the results and code presented in this project are available at: $\tt{https://github.com/benelson/hjs\_with\_stan}$.
This repository includes results for both the NASA Exoplanet Archive and \exoplanetsorg datasets.

\acknowledgements
We thank Joseph Catanzarite, Adam Dempsey, Sybil Derrible, Sam Hadden, Jason Hwang, Yoram Lithwick, Daniel Stevens, Francesca Valsecchi, and Angie Wolfgang for helpful discussions and contributions.
This research has made use of the NASA Exoplanet Archive, which is operated by the California Institute of Technology, under contract with the National Aeronautics and Space Administration under the Exoplanet Exploration Program.
This research has also made use of the Exoplanet Orbit Database and the Exoplanet Data Explorer at \exoplanetsorg.

B.E.N. acknowledges support from the Data Science Initiative at Northwestern University.
E.B.F. acknowledges support from NASA Grant NNX14AN7GG and NNX14AI7GG.
F.A.R. acknowledges support from NASA Grant NNX12AI86G at Northwestern University and from NSF Grant PHY-1066293 through the Aspen Center for Physics.
This research was supported in part through the computational resources and staff contributions provided for the Quest high performance computing facility at Northwestern University which is jointly supported by the Office of the Provost, the Office for Research, and Northwestern University Information Technology.

\bibliography{references}

\clearpage

\end{document}